# ERROR PROBABILITY ANALYSIS OF FREE-SPACE OPTICAL LINKS WITH DIFFERENT CHANNEL MODEL UNDER TURBULENT CONDITION

Bobby Barua[#1], Tanzia Afrin Haque[#2] and Md. Rezwan Islam[#2]

[#1] Assistant Professor, Department of EEE, Ahsanullah University of Science and Technology, Dhaka, Bangladesh

[#2] B.Sc. Engineering Student, Department of EEE, Ahsanullah University of Science and Technology, Dhaka, Bangladesh

`bobby@aust.edu, tanzia_aust@yahoo.com, rezwanislam@yahoo.com,`

## ABSTRACT

*Free space optics (FSO) is a promising solution for the need to very high data rate point-to point communication. FSO communication technology became popular due to its large bandwidth potential, unlicensed spectrum, excellent security and quick and inexpensive setup. Unfortunately, atmospheric turbulence-induced fading is one of the main impairments affecting FSO communications. To design a high performance communication link for the atmospheric FSO channel, it is of great importance to characterize the channel with proper model. In this paper, the modulation format is Q-ary PPM across lasers, with intensity modulation and ideal photodetectors are assumed to investigate the most efficient PDF models for FSO communication under turbulent condition. The performance results are evaluated in terms of symbol error probability (SEP) for different type of channel model and the simulation results confirm the analytical findings.*

## KEYWORDS

## 1. INTRODUCTION

Free Space Optics (FSO) is an optical communication technology that uses light propagating in free space to transmit data between two points [1-2]. Though we assume an LOS path exists between the transmitter and receiver array, the transmitted field from a single laser will propagate through an atmosphere and may experience several effects [3]. First, electromagnetic scattering from water vapor and other molecules causes a redirection of the optical energy, with corresponding loss of signal power at the receiver. Normally, this is only a significant effect if the water vapor content (and drop size) becomes large, or if substantial haze conditions exist.

A second phenomenon is refraction on a more macroscopic scale. Here, small regions of density in homogeneity in the atmosphere, due to pressure and/or temperature gradients, create a non uniform index of refraction throughout the medium. This is especially prominent on optical links parallel to and near the ground. Even though these regions can be treated as lossless, the aggregate field received at some point in the plane of the PDs becomes a random variable. This field strength is a function of space and also time, due to assumed turbulence of the medium. Obviously, the assumption of independence may not be valid, depending upon the spacing of the devices, and on the nature of the fading [4]. The atmospheric turbulence, also called scintillation, one of the most important factor that cause random fluctuations in both the amplitude and the phase of the received signal, what we call channel fading [5]. This turbulence is caused by fluctuations in the refractive index of the medium as the latter experiences temperature gradients due to solar heating and wind. In fact, aerosol scattering effects caused by rain, snow, and fog, can reduce the link range due to the propagation loss in non-clear atmosphere [6]. Even in clear sky conditions, inhomogeneities in the temperature and pressure of the atmosphere caused by solar heating and wind, lead to the variations of the air refractive index along the transmission path [7]. This leads to an increase in the link error probability. A comprehensive survey of optical-propagation effects are found in [8].

To characterize the FSO channel from a communication theory perspective, it is useful to give a statistical representation of the scintillation [9-10]. The reliability of the communication link can be determined if we use a good probabilistic model for the turbulence [11]. Several models exist for the aggregate amplitude distribution, though none is universally accepted, since the atmospheric conditions obviously matter. Most prominent among the models are the Rayleigh, log-normal, gamma-gamma and the negative exponential distribution model.

In this paper, we propose an analytical approach to evaluate the performance of FSO communication system under different type of channel model with Q-ary PPM as modulation technique. The performance results are evaluated in terms of symbol error probability (SEP) assuming that p.i.n. photodiodes are used, and the channel is modeled using Rayleigh, log-normal, gamma-gamma and the negative exponential model.

## 2. SYSTEM MODEL

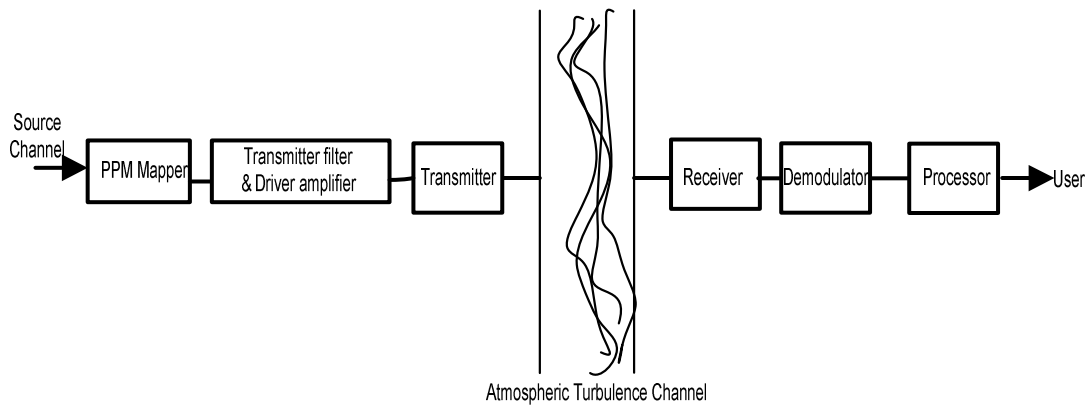

**Fig. 1.** The block diagram of Q-ary PPM system under turbulent condition

Fig. 1 represent the block diagram of QPPM system which transmits **$L=log_2 Q$** bits per symbol, providing high power efficiency. In the transmitter, the signals are described by the binary data bits are converted into a stream of pulses corresponding to QPPM symbol described below, and sent to the laser.

The signals are described by the waveforms

$$\begin{aligned} s_0(t) &= A = \sqrt{2P},\ 0 \leq t \leq T_s/4 & \text{'0 0'} \\ s_1(t) &= A = \sqrt{2P},\ T_s/4 \leq t \leq T_s/2 & \text{'0 1'} \\ s_2(t) &= A = \sqrt{2P},\ T_s/2 \leq t \leq 3T_s/4 & \text{'1 0'} \\ s_3(t) &= A = \sqrt{2P},\ 3T_s/4 \leq t \leq T_s & \text{'0 1'} \end{aligned} \tag{1}$$

At the receiver, after signal detection (the demodulator block) and de-interleaving, channel decoding is performed. The demodulation is performed based on the received signal light intensity. The electrical signal after the optical/electrical conversion is:

$$r_e = \eta(I + I_a) + n \tag{2}$$

where $I$ is the received signal light intensity, $I_a$ is the remaining ambient light intensity after frequency and spatial filtering [9], and     is the optical/electrical conversion efficiency. Also, $n$ is the sum of thermal, dark, and shot noise. We assume that the ambient light can be almost perfectly cancelled [10]. So, after the cancellation of the ambient light, the received signal before demodulation is:

$$r = \eta I + n \tag{3}$$

We suppose that the receiver is thermal noise limited, and consider $n$ as a zero-mean Gaussian additive noise of variance $\sigma_n^2$, independent of the signal $I$. Let $T_s$ denote the symbol duration and $N_0$ the noise unilateral power spectral density. Taking into account the low-pass filtering of bandwidth $1/2T_s$ after photo-detection, the noise variance equals $\sigma_n^2 = N_0/2T_s$. We further consider the received signal intensity $I$ as the product of $I_0$, the emitted light intensity, and $h$, the channel atmospheric turbulence with the PDF given in (1):

$$I = hI_0 \tag{4}$$

## 3. CHANNEL MODELING

A commonly used turbulence model assumes that the variations of the medium can be understood as individual cells of air or eddies of different diameters and refractive indices. In the context of geometrical optics, these eddies may be thought of as lenses that randomly refract the optical wave front, producing a distorted intensity profile at the receiver of a communication system. The most widely accepted theory of turbulence is due to Kolmogorov [12]. This theory assumes that kinetic energy from large turbulent eddies, characterized by the outer scale $L_0$, is transferred without loss to eddies of decreasing size down to sizes of a few millimeters characterized by the inner scale $l_0$. The inner scale represents the cell size at which energy is dissipated by viscosity. The refractive index varies randomly across the different turbulent eddies and causes phase and amplitude variations to the wave front. Turbulence can also cause the random drifts of optical beams–a phenomenon usually referred to as wandering – and can induce beam focusing. The reliability of the communication link can be determined if we use a good probabilistic model for the turbulence. To design a high-performance communication link for the atmospheric free-space optical (FSO) channel, it is of great importance to characterize the channel. Several probability density functions (PDFs) have been proposed for the intensity variations at the receiver of an optical link. The atmospheric turbulence impairs the performance of an FSO link by causing the received optical signal to vary randomly thus giving rise to signal fading. The fading strength depends on the link length, the wavelength of the optical radiation and the refractive index structure parameter $C_n^2$ of the channel. This model is mathematically tractable and it is characterized by the Rytov variance $\sigma_R^2$. The turbulence induced fading is termed weak when $\sigma_R^2 < 1$ and this defines the limit of validity of the model.

$$\sigma_R^2 = 1.23 C_n^2 k^{7/6} L^{11/6} \quad 6$$

(5)

$k = 2\pi/\lambda$ is the optical wave number, $L$ is propagation distance, and $C_n^2$ is the refractive index structure parameter, which we assume to be constant for horizontal paths.

**a. Rayleigh Distribution**

The Rayleigh model is used to describe the channel gain. The scintillation index for the Rayleigh situation is 1. The density function of Rayleigh is more concentrated at low(deeply faded) values.

The PDF for Rayleigh distribution is

$$f(I) = \frac{I}{\sigma_R^2} \exp\left\{-\frac{I}{2\sigma_R^2}\right\}, I \geq 0$$

(6)

**b. Lognormal Distribution**

The log-normal models assumes the log intensity $l$ of the laser light traversing the turbulent atmosphere to be normally distributed with a mean value of $-\sigma_l^2/2$. Thus the probability density function of the received irradiance is given by [11, 12]:

$$f(I) = \frac{1}{(2\pi\sigma_R^2)^{\frac{1}{2}} I} \exp\left\{-\frac{(\ln(I/I_0) + \sigma_R^2/2)^2}{2\sigma_R^2}\right\}, I \geq 0$$

(7)

Where

$I$ represents the irradiance at the receiver
$I_o$ is the signal irradiance without scintillation.

**c. GammaGamma Distribution**

Al-Habash et al. [10] proposed a statistical model that factorizes the irradiance as the product of two independent random processes each with a Gamma PDF. The PDF of the intensity fluctuation is given by [10]

$$f(I) = \frac{2(\alpha\beta)^{(\alpha+\beta)/2}}{\Gamma(\alpha)\Gamma(\beta)} I^{\frac{(\alpha+\beta)}{2}-1} K_{(\alpha-\beta)}(2\sqrt{\alpha\beta I}), I > 0 \tag{8}$$

*I* is the signal intensity, $\Gamma(.)$ is the gamma function, and $K_{\tilde{\alpha}\beta}$ is the modified Bessel function of the second kind and order $\tilde{\alpha}\beta$. $\alpha$ and $\beta$ are PDF parameters describing the scintillation experienced by plane waves, and in the case of zero-inner scale are given by [8]

$$\alpha = \frac{1}{\exp\left[\frac{0.49\sigma_R^2}{(1+1.11\sigma_R^{12/5})^{7/6}}\right] - 1} \tag{9}$$

$$\beta = \frac{1}{\exp\left[\frac{0.51\sigma_R^2}{(1+0.69\sigma_R^{12/5})^{5/6}}\right] - 1} \tag{10}$$

**d. The Negative Exponential Distribution**

In the limit of strong irradiance fluctuations (i.e. in saturation regime and beyond) where the link length spans several kilometers, the number of independent scatterings becomes large . This saturation regime is also called the fully developed speckle regime. The amplitude fluctuation of the field traversing the turbulent medium in this situation is generally believed and experimentally verified to obey the Rayleigh distribution implying negative exponential statistics for the irradiance. That is:

$$f(I) = \frac{1}{I_0} \exp\left(\frac{-I}{I_0}\right), I \geq 0 \tag{11}$$

Where $E[I] = I_o$ is the mean received irradiance. During the saturation regime, the value of the scintillation index, *S.I→1.*

## 4. THEORETICAL ANALYSIS

First, consider the case of negligible background radiation and equal-gain links i.e., $A_{nm}$=1 almost surely, n=1,………..,N, m=1,…….M with no loss of generality, assume that each laser sends energy in slot 1. The only possibility for decision error is that each detector registers zero counts in time slot 1, since the other slots register zero counts by assumption ($n_b$=0) By the Poisson property and independence, we have SEP

$$P_s = \frac{Q-1}{Q}\left[e^{-\frac{M\eta(\frac{P_r}{M})T_s}{hfQ}}\right]^N = \frac{Q-1}{Q}e^{-\frac{\eta E_s N}{hf}} \tag{12}$$

Now for the fading system the probability of zero count in slot 1 at detector n is

$$P[Z_{n1} = 0 \mid slot1, A] = e^{-\sum_{m=1}^{M} a_{nm}^2 \left(\frac{\eta P_r}{Mhf}\right)\left(\frac{T_s}{Q}\right)} \tag{13}$$

If the path gains are independently distributed and identical, the average symbol error is given by [1],

$$P_s = \int P_{s|A} f_A(a) da = \frac{Q-1}{Q}\left\{\left[\int e^{-\frac{a^2\eta\left(\frac{E_s}{M}\right)}{hf}} f_A(a) da\right]^N\right\}^M \tag{14}$$

In case of gamma-gamma fading, the average symbol error becomes

$$P_s = \int P_{s|A} f(I) dI = \frac{Q-1}{Q}\left\{\left[\int e^{-\frac{a^2\eta\left(\frac{E_s}{M}\right)}{hf}} f(I) dI\right]^N\right\}^M \tag{15}$$

## 5. RESULTS AND DISCUSSION

Following the analytical approach presented in section 4 is simulated using Matlab. Here we evaluate the symbol error probability of a FSO link with Q-ary PPM and direct detection scheme under different channel model. In the simulation, the pulse shaping function is assumed to be rectangular and the amplitude {$ac_1$, $ac_2$} of each subcarrier signal is obtained from 1/2ξ. The simulation parameters are given in Table I.

**TABLE I**

| Parameters | Values |
|---|---|
| Data rate | 100 Mbps |
| Modulation type | Q-PPM |
| Sampling frequency | 20 MHz |
| Laser wavelength λ | 850nm |
| PIN photodetector responsivity R | 2 |
| Optical modulation index ξ | 1 |
| Scintillation index | 1 |
| Symbol Energy $E_s$ | $10^{-16}$ Joules |

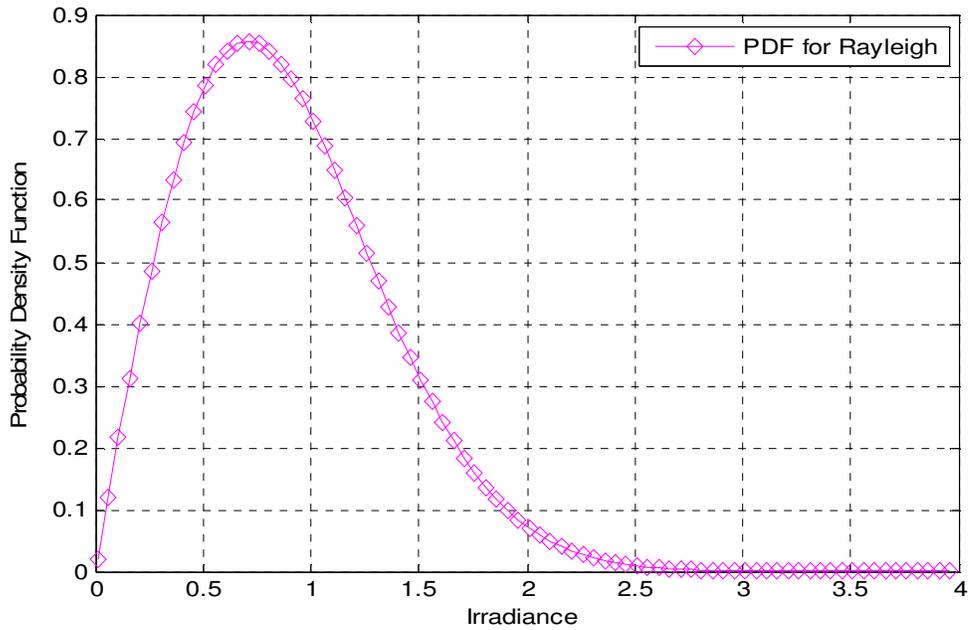

**Fig 2:** Plot of probability density function vs. irradiance for Rayleigh Model.

The plot of the probability density functions for Rayleigh case with typical value of scintillation index (S.I) and turbulence strength is shown in Fig. 2.

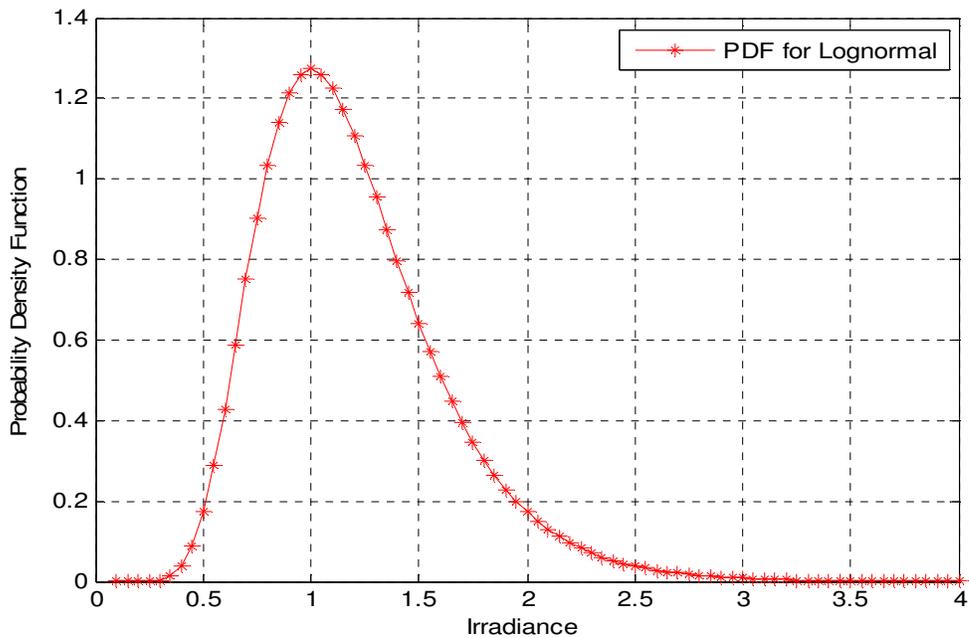

**Fig 3:** Plot of probability density function vs. irradiance for Lognormal Model.

The plot of Fig. 3 shows the probability density functions vs. irradiance for lognormal case with typical value of scintillation index (S.I) and turbulence strength.

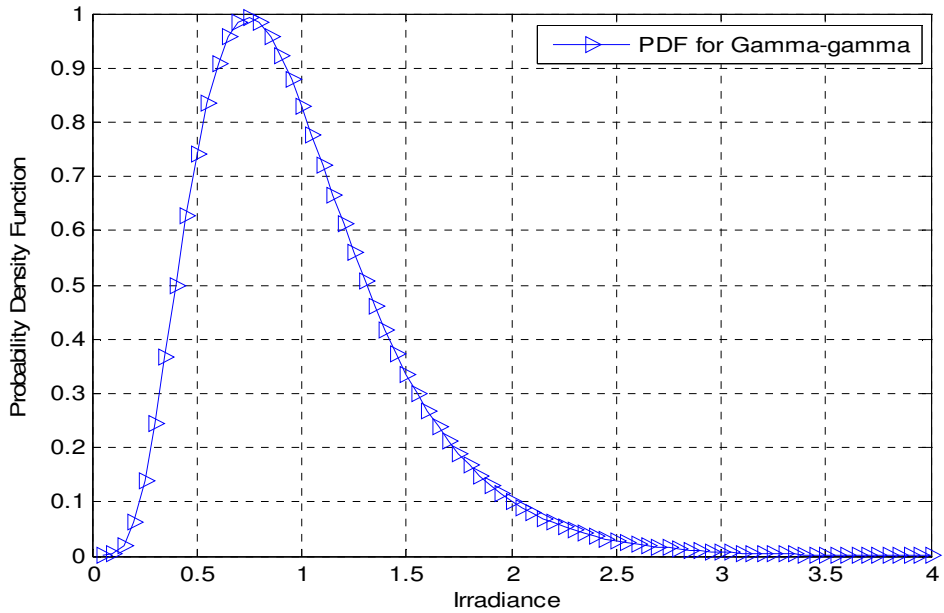

**Fig 4:** Plot of probability density function vs. irradiance for Gamma-gamma Model.

The plot of Fig. 4 shows the probability density functions with respect to the irradiance for gamma-gamma model. . From the figure it is clear that, the probability density function is maximum with the irradiance value of 0.7 for gamma-gamma model.

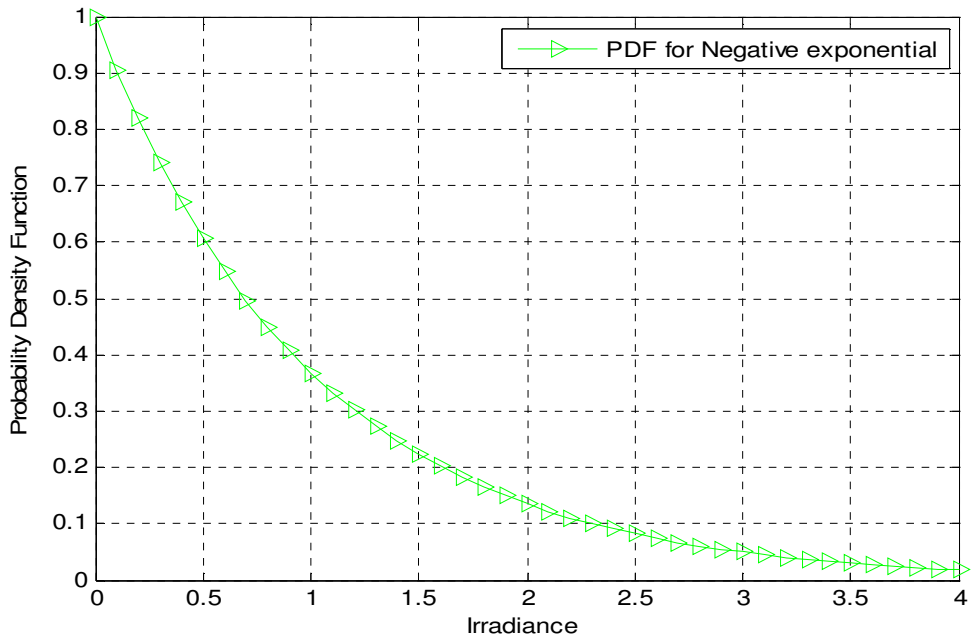

**Fig 5:** Plot of probability density function vs. irradiance for Negative Exponential Model.

Fig. 5 shows the plot of the probability density function for negative exponential model. From the figure it is clear that the optimum value of probability density function is found at negative region. The plots of probability density function for different type of channel models are shown in Fig. 6. From the overall analysis it is clear that the probability density function is maximum for lognormal model at the irradiance value of 1 with typical value of scintillation index (S.I) and turbulence strength.

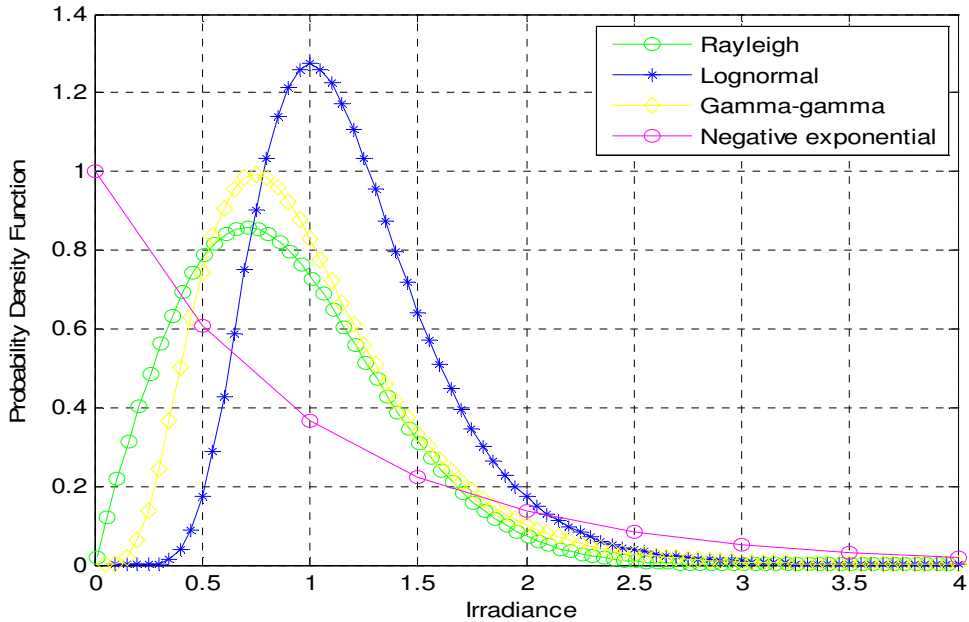

**Fig6:** Plots of probability density function vs. irradiance for different channel model.

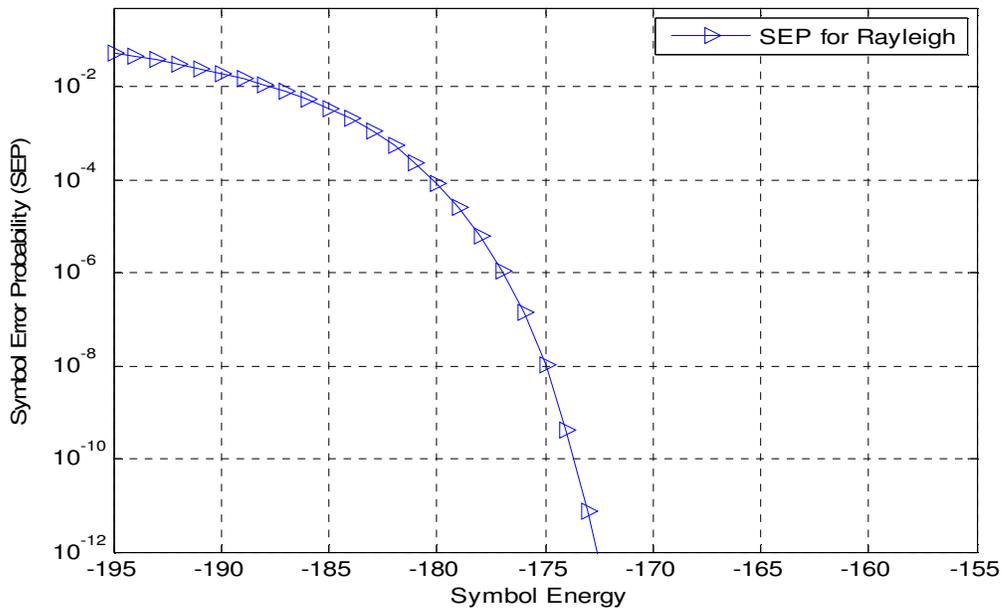

**Fig-7:** Symbol-error probability vs. symbol energy for rayleigh channel model

International Journal of Computer Science & Information Technology (IJCSIT) Vol 4, No 1, Feb 2012

Fig. 7 shows the SEP versus symbol energy for rayleigh model with Q-ary scheme. Rayleigh fading emerges from a scattering model that views the composite field as produced by a large number of nondominating scatterers, each contributing random optical phase upon arrival at the detector. The symbol energy due to background light is set to -170 dBJ.

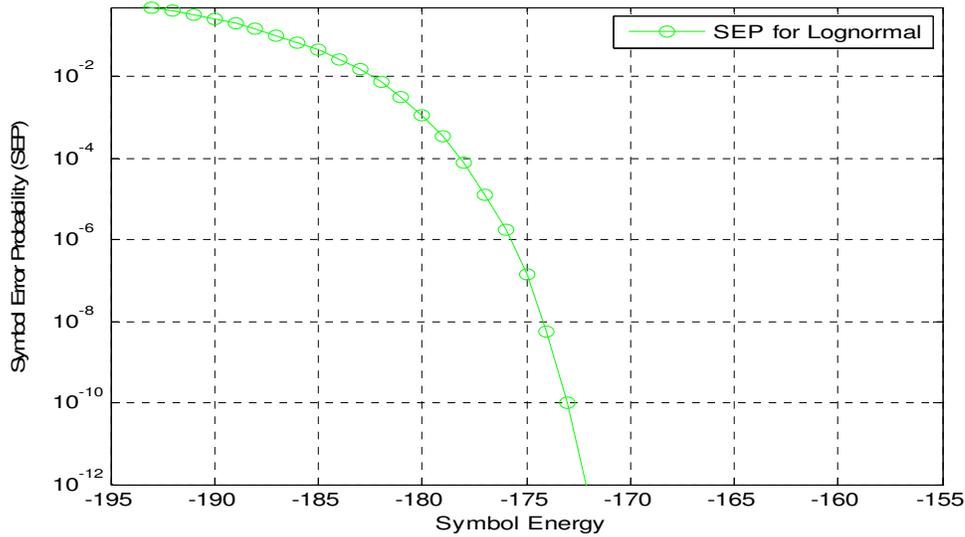

**Fig-8:** Symbol-error probability vs. symbol energy for lognormal channel model

The plot of symbol error probability versus symbol energy for lognormal model is shown in Fig.-8 under a background radiation with energy $10^{-17}$ joules - which is close to the background level we calculate in section 4 fixed for binary PPM and QPPM.

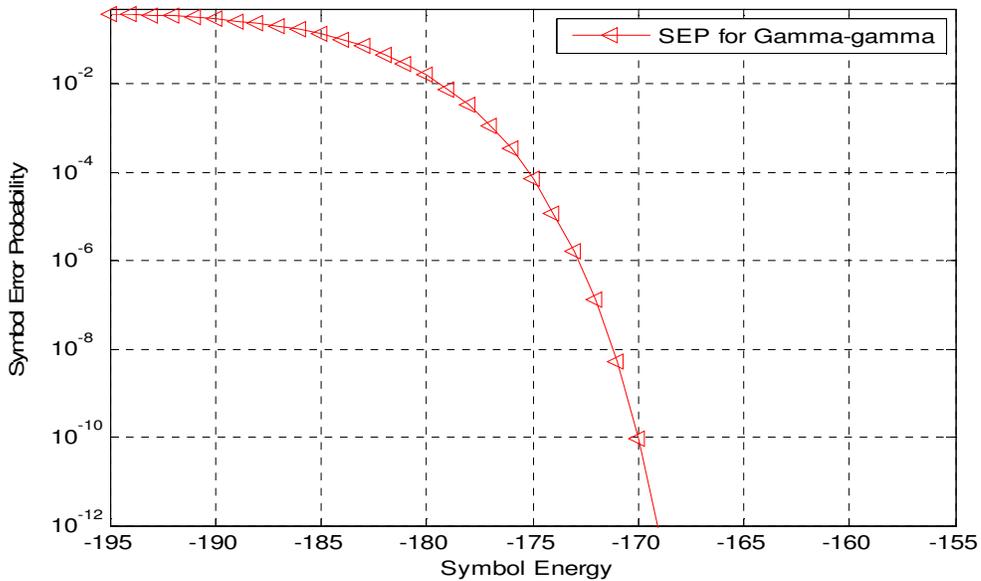

**Fig-9:** Symbol-error probability vs. symbol energy for gamma-gamma channel model

254

The SEP versus symbol energy for gamma-gamma model with Q-ary scheme is shown in Fig. 9. In particular, notice the gamma-gamma model has a much higher density in the high amplitude region, leading to a more severe impact on system performance. The symbol energy due to background light is also set to -170 dBJ. Fig.-10 shows the plot of symbol error probability versus symbol energy for negative exponential model.

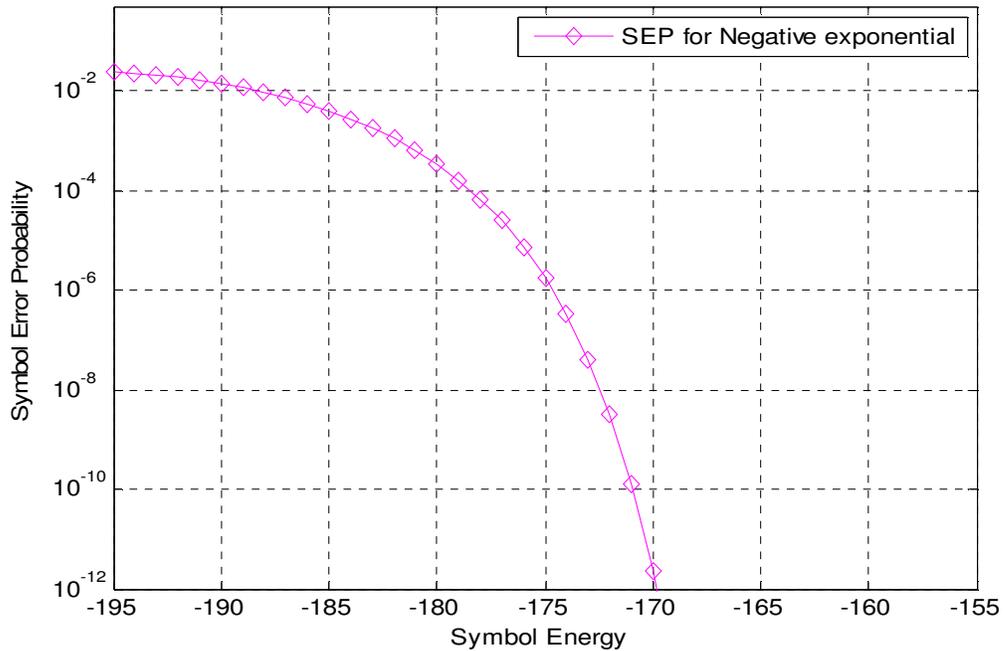

**Fig-10:** Symbol-error probability vs. symbol energy for negative exponential channel model

The plots of symbol error probability versus symbol energy for different type of channel models are shown in the Fig. 11. From the analysis of the plots it is clear that the symbol energy at SEP $10^{-12}$ is almost similar for both rayleigh and gamma gamma case. But the Rayleigh case is the limiting version of a more general Rician model, also advocated as a fading model similar to the log-normal case. Again the symbol energy is almost similar for lognormal and negative exponential model at SEP $10^{-12}$. . From the overall analysis it is found that the gamma-gamma model performs better at any value of symbol energy.

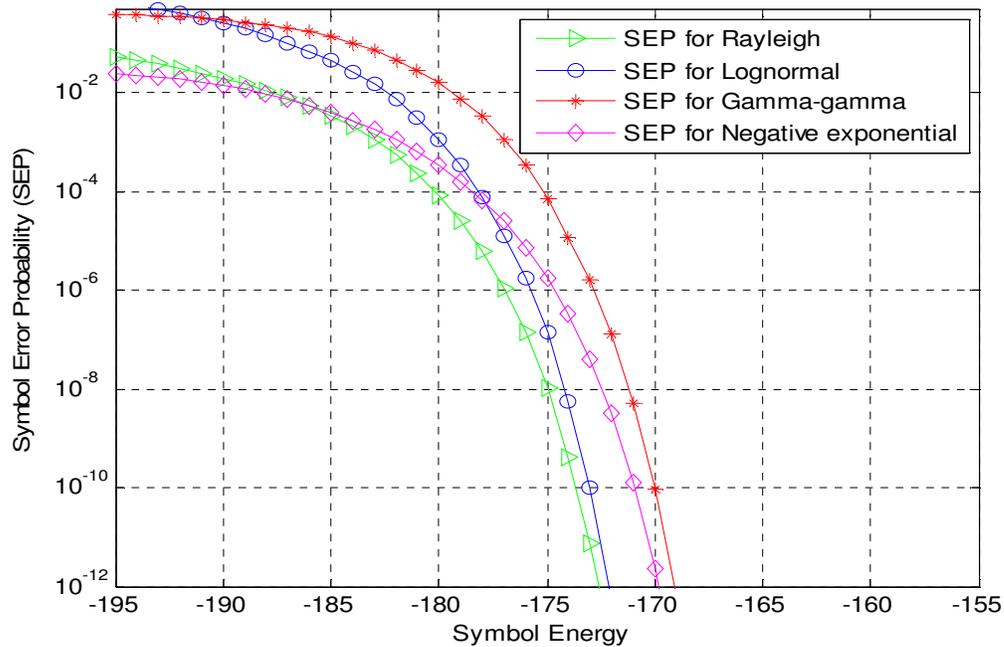

**Fig-11:** Symbol-error probability vs. symbol energy for different channel model.

## 6. CONCLUSIONS

A detailed analytical approach is presented to evaluate the symbol error probability under different type of channel models used in FSO communication for the turbulence channel models such as the Rayleigh, Log-normal, Gamma–gamma and Negative exponential model distribution are valid from weak to strong turbulence regime. In particular, notice the Rayleigh model has a much higher density in the low amplitude region, leading to a more severe impact on system performance and also advocated as a fading model similar to the log-normal case under consideration of S.I→1. It should be noted that the Gamma–gamma model performs better for all regimes from weak to strong turbulence region. The negative exponential model is also valid for the same limit of Gamma-gamma model but the optimum value is occurred at negative region. So finally our decision is to prefer gamma-gamma model under weak to strong turbulence regime as channel model for FSO communication.

International Journal of Computer Science & Information Technology (IJCSIT) Vol 4, No 1, Feb 2012

3. V. I. Tatarskii, *Wave Propagation in a Turbulent Medium* (Dover Publications Inc., 1968). New York.

4. J. Strohbehn, Ed. "*Laser Beam Propagation in the Atmosphere*" New York: Springer, 1978.

5. L. C. Andrews, R. L. Phillips, C. Y. Hopen, M. A. Al-Habash, "*Theory of optical scintillation*," J. Opt. Soc. Am. A 16, 1417–1429 (1999).

6. G. Ochse, *Optical Detection Theory for Laser Applications*. NewYork: Wiley- Interscience, 2002.

7. L.C. Andrews and R. L. Phillips, *Laser Beam Propagation through Random Media* (SPIE Press, Bellingham,Washington, 2005), 2nd ed.

8. M. A. Al-Habash, L. C. Andrews, and R. L. Phillips, "Mathematical model for the irradiance probability density function of a laser beam propagating through turbulent media," Opt. Eng. 40, 1554–1562 (2001).

9. R. M. Gagliardi and S. Karp, *Optical Communications* (John Wiley & Sons, 1995), 2nd ed.

10. H. Willebrand and B. S. Ghuman, Free Space Optics: Enabling Optical Connectivity in Todays Network. Indianapolis, IN: SAMS, 2002.

11. X. Zhu and J. Kahn, "Free-space optical communication through atomospheric turbulence channels," IEEE Trans. Commun. 50, 1293–1330 (2002).

12. A. N. Kolomogrov, "On the Shannon theory of information transmission in the case of continuous signals," IRE Transactions on Information Theory, vol. 2, no. 4, pp. 102–108, December 1956.

**Author**

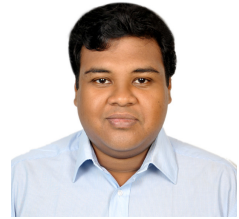

**Bobby Barua** received the B.Sc. and M.Sc. degrees in electrical and electronic engineering from Ahsanullah University of Science and Technology and Bangladesh University of Engineering and Technology respectively. He is currently Assistant Professor of Electrical and Electronic Engineering at Ahsanullah University of Science and Technology, Dhaka, Bangladesh. His research interests are in applications of information theory and coding to modern communication systems, specifically digital modulation and coding techniques for satellite channels, wireless networks, spread-spectrum technology, and space–time coding for multipath channels.

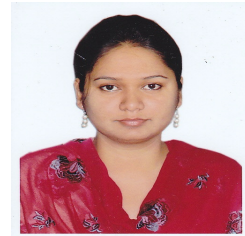

**Tanzia Afrin Haque** received the Bachelor's Degree in Electrical & Electronics Engineering from Ahsanullah University of Science & Technology, Dhaka, Bangladesh in 2012. She is currently trying to go abroad for higher study. Her area of interest includes free-space optical communication & media communication. She has several publications in free space optical communication.

**Md. Rezwan Islam** received his Bachelor's Degree in Electrical &Electronics Engineering from Ahsanullah University of Science & Technology Dhaka, Bangladesh in 2012. He is currently trying to go abroad for higher study. He has one international journal & one conference paper based on free-space optical communication. His field of interest includes free-space optical communication & different types of turbulence in communication medium.

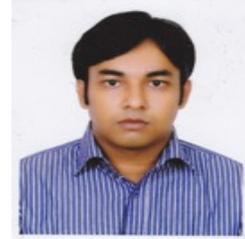